\documentclass[aps,twocolumn,showpacs,superscriptaddress,amssymb]{revtex4-1}
\date{\today}
\usepackage{graphicx,threeparttable,amssymb}
\usepackage{dcolumn}
\usepackage{bm}
\usepackage{color}

\begin{document}

\title{Mirror energy difference and the structure of loosely bound proton-rich nuclei around $A=20$}

\author{Cenxi Yuan}
\affiliation{Sino-French Institute of Nuclear Engineering and
Technology, Sun Yat-Sen University, Zhuhai 519082, China}
\affiliation{State Key Laboratory of Nuclear Physics and Technology,
School of Physics, Peking University, Beijing 100871, China}
\author{Chong Qi}
\affiliation{Royal Institute of Technology, AlbaNova University
center, SE-10691 Stockholm, Sweden}
\author{Furong Xu}
\email{frxu@pku.edu.cn} \affiliation{State Key Laboratory of Nuclear
Physics and Technology, School of Physics, Peking University,
Beijing 100871, China}
\author{Toshio Suzuki}
\affiliation{Department of Physics, College of Humanities and
Sciences, Nihon University, Sakurajosui 3, Setagaya-ku, Tokyo
156-8550, Japan} \affiliation{National Astronomical Observatory of
Japan, Mitaka, Tokyo 181-8588, Japan }
\author{Takaharu Otsuka}
\affiliation{Department of Physics, University of Tokyo, Hongo,
Bunkyo-ku, Tokyo 113-0033, Japan} \affiliation{Center for Nuclear
Study, University of Tokyo, Hongo, Bunkyo-ku, Tokyo 113-0033, Japan}
\affiliation{National Superconducting Cyclotron Laboratory, Michigan
State University, East Lansing, Michigan, 48824, USA}

\begin{abstract}

The properties of loosely bound proton-rich nuclei around $A=20$ are
investigated within the framework of nuclear shell model. In these
nuclei, the strength of the effective interactions involving the
loosely bound proton $s_{1/2}$ orbit are significantly reduced in
comparison with those in their mirror nuclei. We evaluate the
reduction of the effective interaction by calculating the
monopole-based-universal interaction ($V_{MU}$) in the Woods-Saxon
basis. The shell-model Hamiltonian in the $sd$ shell, such as USD,
can thus be modified to reproduce the binding energies and energy
levels of the weakly bound proton-rich nuclei around $A=20$. The
effect of the reduction of the effective interaction on the
structure and decay properties of these nuclei is also discussed.

\end{abstract}


\maketitle

\section{\label{sec:level1}Introduction}
The study of proton-rich nuclei plays an important role in the
understanding of a variety of nuclear astrophysical
processes~\cite{grawe2007}, such as the $^{17}$F$(p,\gamma)^{18}$Ne
reaction in stellar explosions~\cite{bardayan1999}. The excitation
spectra of proton-rich nuclei are similar to those in their mirror
partners because the strong Nucleon-Nucleon (NN) interaction is
almost charge independent and the influence of the Coulomb
interaction on the excitation spectra is relatively
small~\cite{bohr1969,brown2001}.

For heavier nuclei in the $fp$ shell, the energy difference between
mirror states (MED) are rather small (usually only around $0.1$~MeV)
~\cite{warner2006,bentley2007,qi2008}. However, for light nuclei,
the MED can be one order of magnitude larger. For example, the
energy of the $1/2^{+}_{1}$ state in $^{13}$N is $0.72$~MeV lower
than that in $^{13}$C~\cite{nndc}. This shift in energy is related
to the loosely bound nature of the proton $1s_{1/2}$
orbit~\cite{thomas,ehrman}. Since there is no centrifugal barrier,
the radial wave function of the $1s_{1/2}$ orbital extends into a
much larger space than those of other neighboring orbitals. Thus the
Coulomb energy of the weakly bound $1s_{1/2}$ orbit,
$\langle1s_{1/2}|V_{C}|1s_{1/2}\rangle$, is less repulsive than
those of other orbits and forms the shift of the $1/2^{+}_{1}$ state
from $^{13}$C to $^{13}$N. Due to the Coulomb force and the
isospin-nonconserving term of the nuclear force, the residual
interaction $V^{pp}$ in proton-rich nuclei are typically a few
percent weaker than the corresponding $V^{nn}$ in their mirror
nuclei~\cite{ormand1989}. However, for the $V^{pp}$ related to the
weakly bound $1s_{1/2}$ orbit, the ratio $V^{pp}/V^{nn}$ can be as
small as $0.7$, which can be deduced from observed data in nuclei
around $^{16}$O~\cite{ogawa1999}.

In nuclei around $A=20$, where the $1s_{1/2}$ orbit plays an
important role, the excitation energies of some states in
proton-rich nuclei show large discrepancy when compared to their
mirror states. For example, the astrophysically important
$3_{1}^{+}$ state in $^{18}$Ne is lower than the corresponding state
in $^{18}$O by about $800$ keV~\cite{bardayan1999}. This $3_{1}^{+}$
state in $^{18}$Ne is above the proton separation threshold and
quasi-bound due to the Coulomb barrier. It is expected that the
following two aspects can be important in contributing to the
difference between these mirror nuclei: the shift of the single
particle energies and the reduction of the proton-proton residual
interaction.

There are several well established shell-model Hamiltonians in the
$sd$ shell, such as USD~\cite{usd}, USDA~\cite{usdab} and
USDB~\cite{usdab}.  These are obtained by fitting to the binding
energies and the excitation energies of the low-lying levels of
nuclei with $N\geq Z$. However, proton-rich nuclei are affected by a
mechanism not incorporated into the USD family even if they are
phenomenologically optimized. The loosely binding effect of the
proton orbitals is not taken into account in the USD family. On
proton-rich side of the $sd$ shell, the proton $d_{5/2}$ and
$s_{1/2}$ orbitals are weakly bound or quasi bound in some nuclei,
while both are deeply bound on the neutron-rich side.

In this paper, we will study the structure and decay properties of
the weakly bound proton-rich nuclei around $A=20$ by using the
nuclear shell model with above effective interactions. It is
expected that the binding energies and excitation spectra of these
proton-rich nuclei can be reproduced by modifying the
single-particle energies and the two-body matrix elements (TBME) of
the existing Hamiltonians. The weakly bound effect is dominated by
the interplay between the spreading of radial wave functions and
finite range properties of nuclear forces. Thus, the reduction
factors of TBME are evaluated with the newly introduced NN
interaction, monopole based universal interaction $(V_{MU})$ which
has explicit dependence on the inter-nucleon distance and has been
shown to be reasonable for basic properties like
monopoles~\cite{otsuka2010}.

 In this work we will
evaluate the reduction effect of the TBME from a phenomenological
point of view. It should be mentioned that the present work can also
be helpful for future microscopic studies with realistic NN
interaction. In particular, in Ref.~\cite{kuo1997}, it is argued
that the core-polarization effect can be dramatically suppressed in
halo nuclei.

In Sec.~\ref{sec:level2}, we evaluate the reduction factors for the
related TBME. The properties of loosely bound proton-rich nuclei
around $A=20$ are discussed in Sec.~\ref{sec:level3}. This work is
summarized in Sec.~\ref{sec:level4}.

\section{\label{sec:level2}Theoretical framework}

\begin{figure}
\includegraphics[scale=0.3]{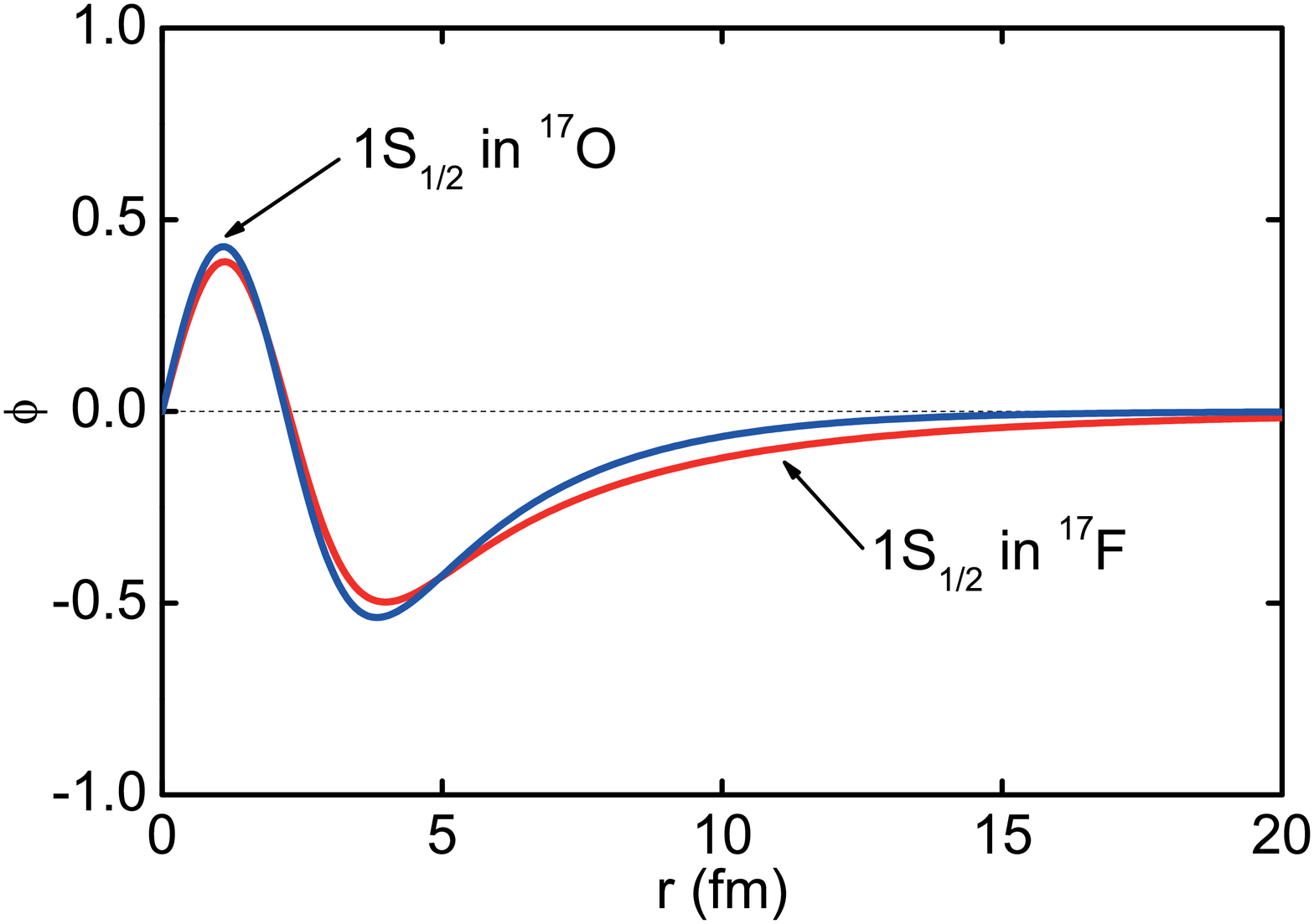}
\caption{\label{wfsd} (Color online) Single-particle wave functions
of the $1s_{1/2}$ orbits in $^{17}$O and $^{17}$F.}
\end{figure}

The radial wave function of the proton $1s_{1/2}$ orbit in loosely
bound proton-rich nuclei extends into a larger coordinate space than
that of the neutron $1s_{1/2}$ orbit in the corresponding mirror
nuclei. As an illustration, in Fig.~\ref{wfsd} we show the
calculated radial wave functions of the valence $1s_{1/2}$ orbits in
nuclei $^{17}$F and $^{17}$O. The calculations are done with the
Woods-Saxon potential with the depth $V_{0}=50.9~(50.2)$ ~MeV for
$^{17}$F ($^{17}$O). The depths are determined by fitting to the
single-particle energies of the $1s_{1/2}$ states, which are
$-0.10$~MeV and $-3.27$~MeV in $^{17}$F and $^{17}$O,
respectively~\cite{nndc}. Here, we assume that these energies can be
set equal to measured one nucleon separation energies. These depths
are close to the one given in Ref.~\cite{bohr1969}. The diffuseness
and radius parameters in the Woods-Saxon potential are chosen to be
$a=0.67$~fm and $R=1.27A^{1/3}$~fm~\cite{heydebasic}, respectively,
where $A$ is the mass number of the nucleus. It can be clearly seen
from Fig. \ref{wfsd} that the $1s_{1/2}$ orbit in $^{17}$F has a
larger space distribution than that in $^{17}$O. Earlier
experimental \cite{f17} and theoretical \cite{Ren98} analyses also
show that the single proton in $1/2^+$ state in $^{17}$F has very
large space distribution. It should be mentioned that, for a
reasonable set of Woods-Saxon parameters, the radial part of
single-particle wave function is not sensitive to the detail of the
potential. We find that the single-particle wave functions of the
$0d_{5/2}$ orbit in $^{17}$F and $^{17}$O are rather similar to each
other. The radial wave functions of the $0d_{3/2}$ orbital in
$^{17}$F and $^{17}$O, both of which are unbound, are calculated
using the code GAMOW \cite{vertse1982} with above Woods-Saxon
parameters. Despite the different outgoing waves because of the
different resonant widths of the $3/2^+$ state in $^{17}$F and
$^{17}$O, the radial wave functions of $0d_{3/2}$ orbit in these two
nuclei are quite similar inside the nuclei. The $0d_{3/2}$ orbit is
also relatively less important compared with $0d_{5/2}$ and
$1s_{1/2}$ orbits in the study of low-lying states of the nuclei
around $A=20$. These are the reasons why only the TBME related to
$1s_{1/2}$ proton need to be modified, which will be discussed
later.


Our shell-model effective Hamiltonian is constructed starting from
the well established USD, USDA and USDB interactions. The USD family
has been determined by fitting to nuclei in the neutron-rich side by
assuming isospin symmetry.

In the present work, the charge symmetry breaking of the NN
interaction is not taken into account because the mirror differences
are mostly caused by the weakly bound protons in the nuclei being
studied as mentioned in the introduction. Calculations with the
charge-dependent Bonn potential \cite{Mac01,qi2008} show that the
effect of the charge dependence in the $sd$ shell nuclei is rather
minor. This is consistent with the result of Ref.~\cite{ormand1989}.
The single-proton energy of the $1s_{1/2}$ orbit, relative to the
one of the $0d_{5/2}$ orbit,  of the shell-model Hamiltonian is
lowered by $0.375$~MeV as compared to the neutron one, by taking
into account the fact that the experimental excitation energy of the
$1/2^{+}_{1}$ state in $^{17}$F is $0.375$~MeV lower than that in
$^{17}$O.

The reduction factor of TBME, $f=\langle ij|V|kl\rangle^{pp}/\langle
ij|V|kl\rangle^{nn}$, is obtained with the Woods-Saxon
single-particle wave function and an effective NN interaction. Here
we use $V_{MU}$~\cite{otsuka2010} plus
 the spin-orbit force from the M3Y
interaction~\cite{m3y1977} as the NN interaction. $V_{MU}$, which
includes the Gaussian type central force and the $\pi+\rho$ bare
tensor force, can explain the shell evolution in a large region of
nuclei~\cite{otsuka2010}. The original $V_{MU}$ parameters can
reproduce well the monopole part of SDPF-M and GXPF1A interactions
in $sd$ and $pf$ regions~\cite{otsuka2010}. The validity of the
$V_{MU}$ in shell-model calculation is examined in the
$psd$~\cite{yuan2012} and $sdpf$~\cite{utsuno2012,suzuki2012}
regions. A similar method was used in Ref.~\cite{ogawa1999} to
evaluate the reduction factor by using the M3Y interaction.

One needs a transformation from relative coordinate to usual
shell-model coordinate in order to obtain TBME $\langle
ij|V|kl\rangle$ from the NN interaction. We expand $\langle
ij|V|kl\rangle_{WS}$ in harmonic oscillator basis. A Woods-Saxon
single-particle wave function, e.g., $|1s_{1/2}\rangle_{WS}$, is
expanded in ten harmonic oscillator single-particle wave functions,
saying $|Ns_{1/2}\rangle_{HO}$~(N is from $0$ to $9$). The harmonic
oscillator wave functions are calculated with the parameter
$\hbar\omega=45A^{-1/3}-25A^{-2/3}$ ($A=18$).

Our calculations thus show that only two-body interactions related
to the proton $1s_{1/2}$ orbit are noticeably modified by
calculations with the Woods-Saxon potential. In Table~\ref{reduce}
we give the reduction factors of five proton-proton TBME involving
the $1s_{1/2}$ orbit. A microscopic study shows a similar magnitude
of reduction factors in weakly bound neutron-rich nuclei with Skyrme
Hartree-Fock basis~\cite{signoracci2011}. The reduction effect of
other TBME is assumed to be much weaker, and is not taken into
account in the following calculations for simplicity.

\begin{table}
\caption{\label{reduce}Calculated reduction factors for the five proton-proton TBME in which  the
$1s_{1/2}$ orbit is involved.}
\begin{ruledtabular}
\begin{tabular}{cc}
  TBME ($\langle ij|V|kl\rangle^{pp}_{JT}$)&Reduction factor\\
\hline $\langle(1s_{1/2})^{2}|V|(1s_{1/2})^{2}\rangle^{pp}_{01}$&0.68\\
$\langle1s_{1/2}0d_{5/2}|V|1s_{1/2}0d_{5/2}\rangle^{pp}_{31}$&0.78\\
$\langle1s_{1/2}0d_{5/2}|V|1s_{1/2}0d_{5/2}\rangle^{pp}_{21}$&0.84\\
$\langle(0d_{5/2})^{2}|V|(1s_{1/2})^{2}\rangle^{pp}_{01}$&0.80\\
$\langle1s_{1/2}0d_{5/2}|V|(0d_{5/2})^{2}\rangle^{pp}_{21}$&0.87
\\
\end{tabular}
\end{ruledtabular}
\end{table}

\begin{figure}
\includegraphics[scale=0.35]{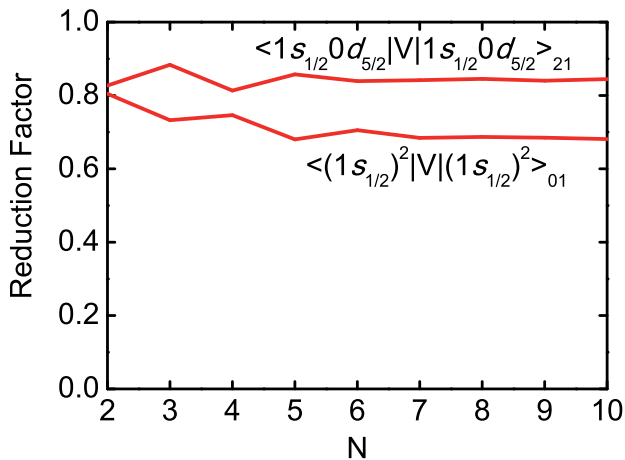}
\caption{\label{nhw} (Color online) Reduction factors as the
function of $N$, which means a Woods-Saxon wave function is expanded
in $N$ harmonic oscillator wave functions}
\end{figure}

We have tested how the reduction factors depend on the mass number
$A$ and the number of harmonic oscillator shells $N$. The reduction
factors are almost independent of $A$ and converge after $N=7$.
Figure~\ref{nhw} gives the convergence of reduction factors for
$\langle(1s_{1/2})^{2}|V|(1s_{1/2})^{2}\rangle^{pp}_{01}$ and
$\langle1s_{1/2}0d_{5/2}|V|1s_{1/2}0d_{5/2}\rangle^{pp}_{21}$ as the
function of the numbers of harmonic oscillator basis which are used
in our expansion of the Woods-Saxon wave function. Here, we evaluate
the reduction factors through $V_{MU}$ because the weakly bound
effect of proton $1s_{1/2}$ orbit is not included in the USD family.
As the USD family performs very well and is used widely in the study
of neutron-rich side of the $sd$ shell, we use the modified USD
family to study the spectroscopic properties of the nuclei being
studied. The Hamiltonians are labeled as USD*, USDA* and USDB* when
the reduction modification is made.

\section{\label{sec:level3}The structure of loosely bound proton-rich nuclei around $A=20$}

\begin{figure}
\includegraphics[scale=0.33]{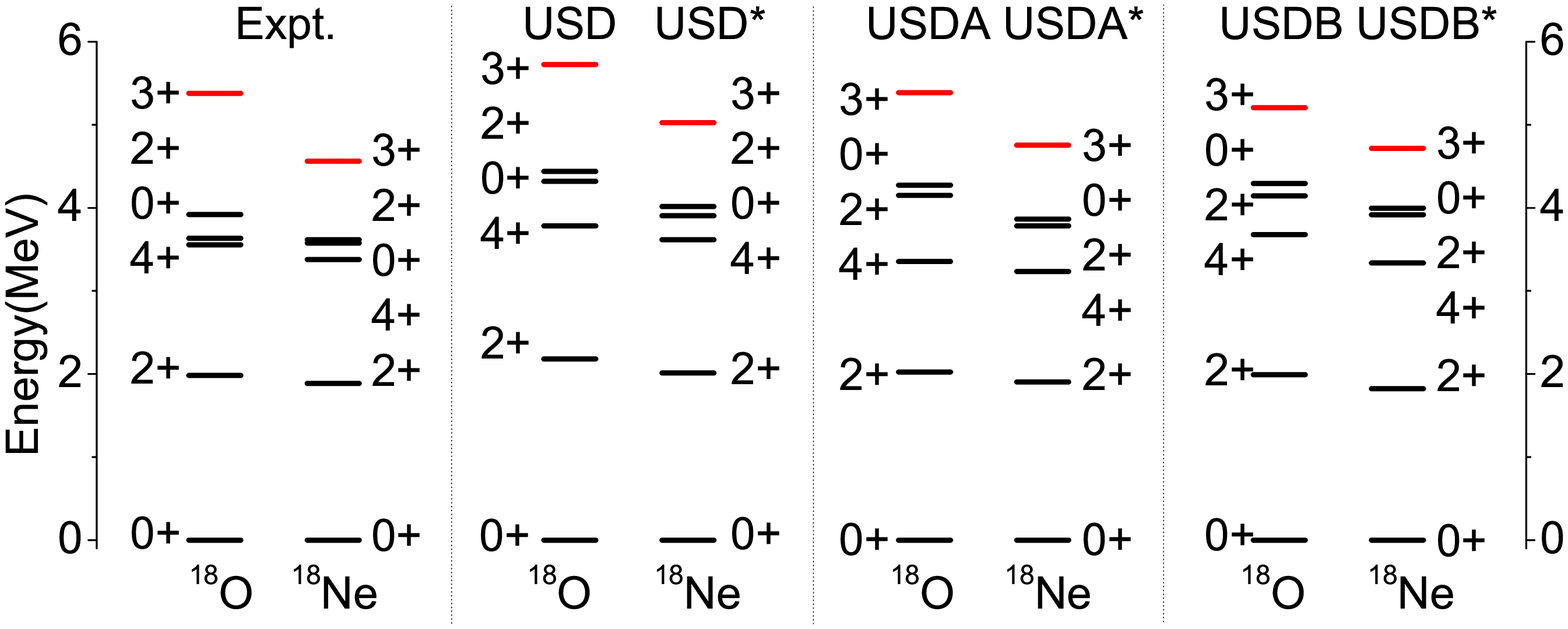}
\caption{\label{18Ne} (Color online) Experimental and calculated
levels of the mirror nuclei $^{18}$O and $^{18}$Ne. USD*, USDA* and
USDB* indicate the calculations with the modified proton-proton TBME
(see Table~\ref{reduce} for exact modification factors and
corresponding text for explanations). Data are from
Ref.~\cite{nndc}.}
\end{figure}

\begin{figure}
\includegraphics[scale=0.33]{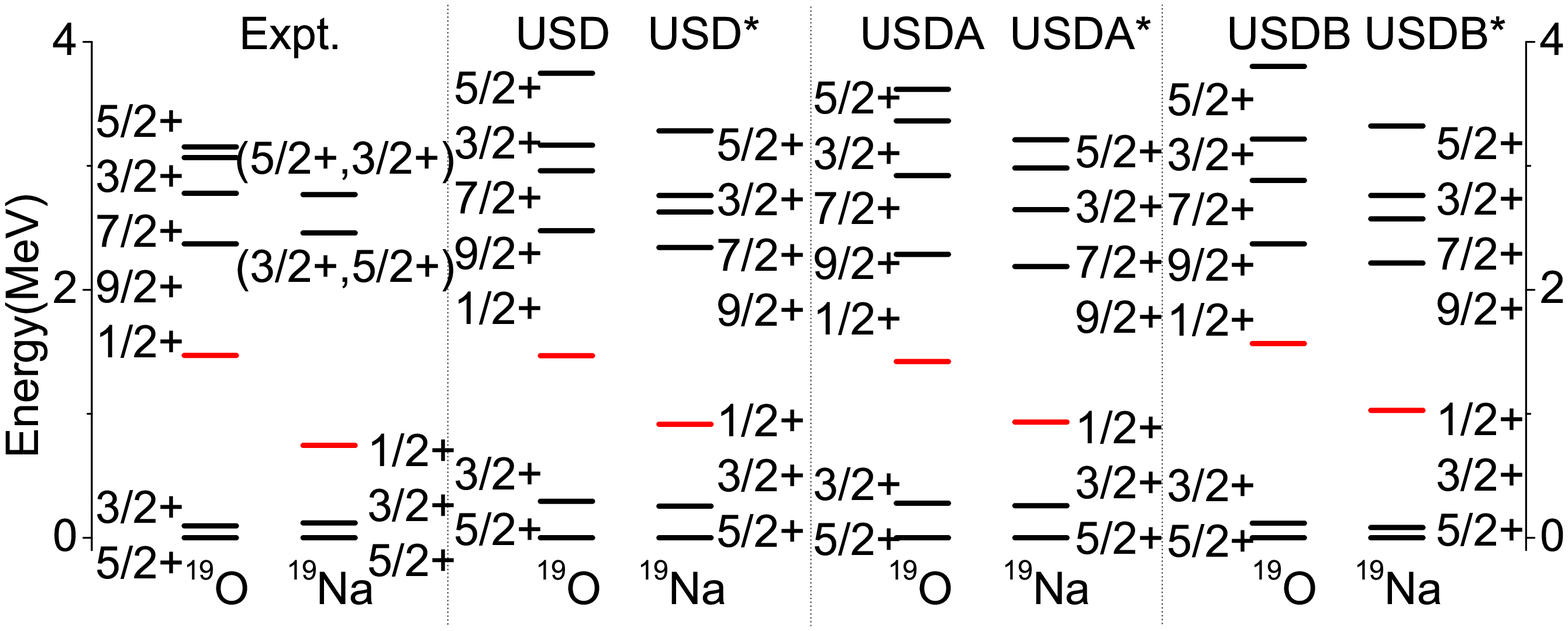}
\caption{\label{19Na} (Color online) Same as Fig. \ref{18Ne} but for
mirror nuclei $^{19}$O and $^{19}$Na. Data are from
Refs.~\cite{nndc,angulo2003,pellegriti2008}.}
\end{figure}

\begin{figure}
\includegraphics[scale=0.33]{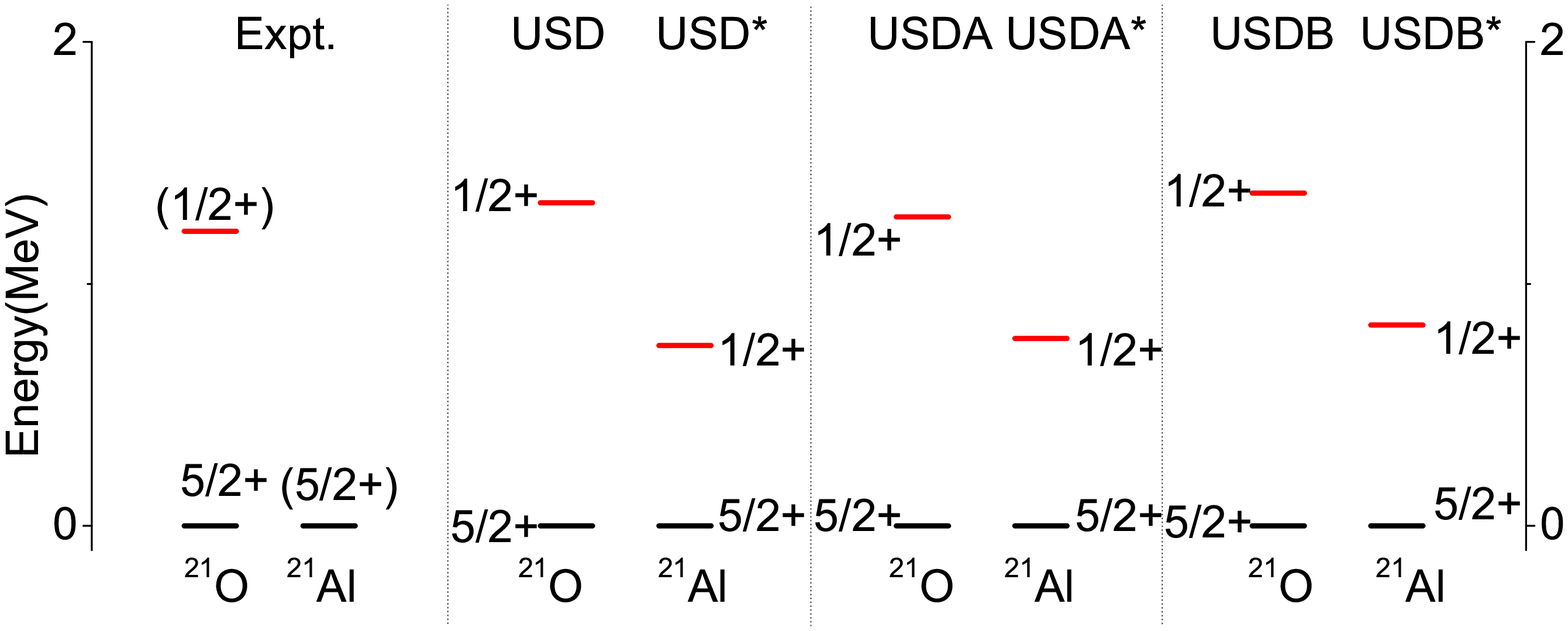}
\caption{\label{21Al} (Color online) Same as Fig. \ref{18Ne} but for
mirror nuclei $^{21}$O and $^{21}$Al. Data are from
Ref.~\cite{nndc}.}
\end{figure}

\begin{figure}
\includegraphics[scale=0.165]{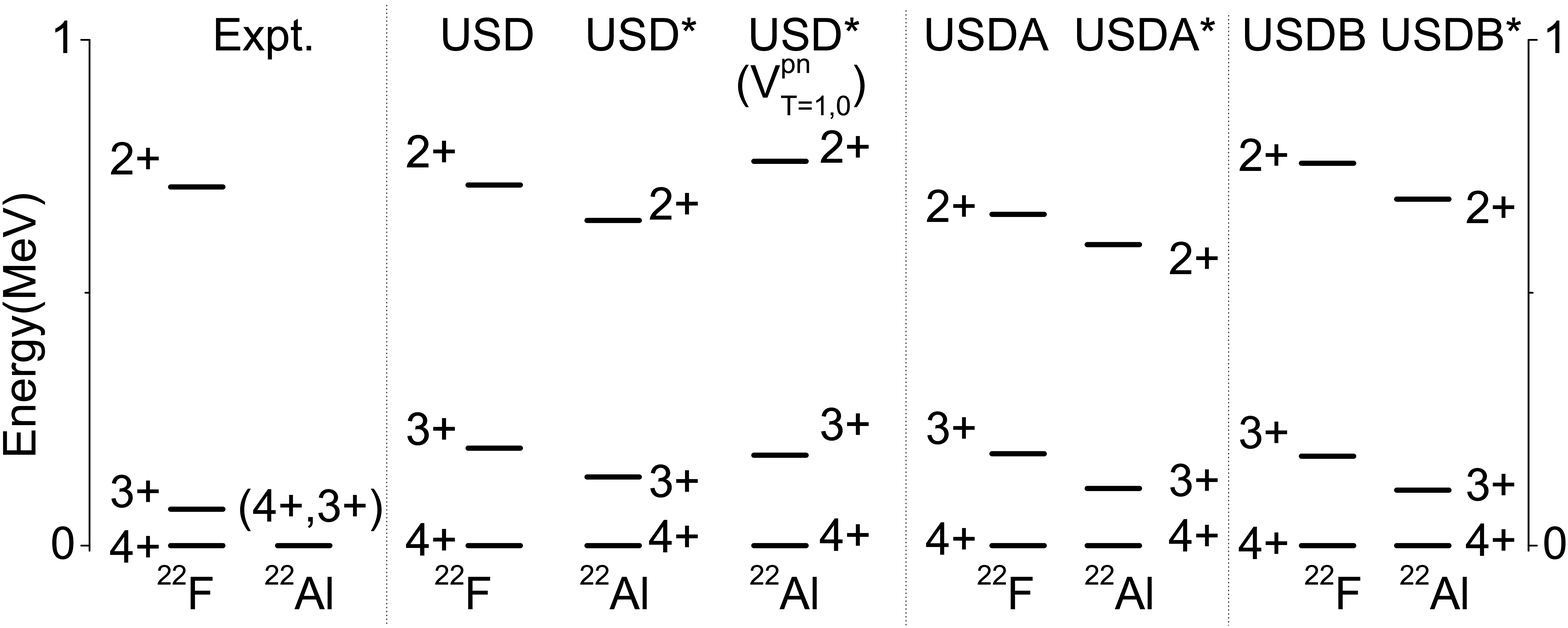}
\caption{\label{22Al} (Color online) Same as Fig. \ref{18Ne} but for
mirror nuclei $^{22}$F and $^{22}$Al. Data are from
Ref.~\cite{nndc}.}
\end{figure}

\begin{figure}
\includegraphics[scale=0.165]{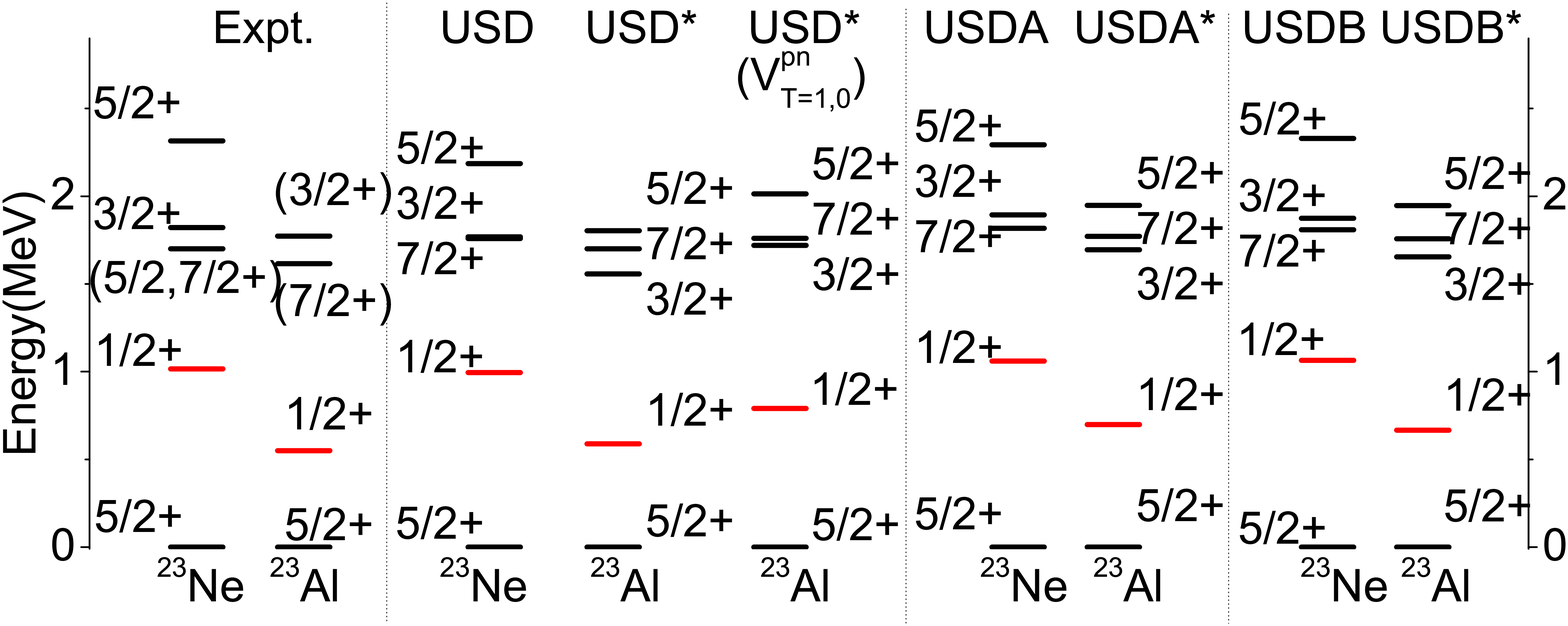}
\caption{\label{23Al} (Color online) Same as Fig. \ref{18Ne} but for
mirror nuclei $^{23}$Ne and $^{23}$Al. Data are from
Refs.~\cite{nndc,gade2008}.}
\end{figure}

\begin{figure}
\includegraphics[scale=0.165]{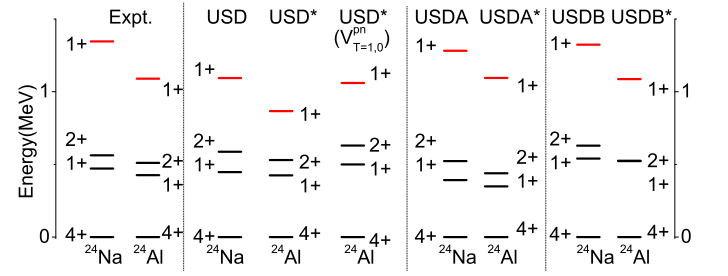}
\caption{\label{24Al} (Color online) Same as Fig. \ref{18Ne} but for
mirror nuclei $^{24}$Na and $^{24}$Al. Data are from
Refs.~\cite{nndc,ichikawa2009}.}
\end{figure}

Calculations are done in the $sd$ shell by employing the shell model
code OXBASH~\cite{oxbash} with the effective Hamiltonians mentioned
above. In the following, we will concentrate on proton drip-line
nuclei $^{18}$Ne, $^{19}$Na, $^{20}$Mg, $^{21-24}$Al and
$^{22-24}$Si, where the proton $1s_{1/2}$ orbital is weakly or quasi
bound. In Refs.~\cite{fortune1,fortune2,fortune3}, the proton-rich
nuclei $^{18,19}$Mg are studied within a Woods-Saxon potential model
by considering the shell-model spectroscopic factors.

In Figs.~\ref{18Ne} to \ref{24Al} we show the comparison between the
experimental and calculated energy levels of $^{18}$Ne, $^{19}$Na,
$^{21-24}$Al and those of their mirror nuclei.
The interaction, USD*($V^{pn}_{T=1,0}$), and related results in Figs.~\ref{22Al} to \ref{24Al} will be discussed later.
The original USD,
USDA, USDB results can be found in Ref.~\cite{oxbash}. It is thus
seen that the MED's of the analogous states can be reproduced very
well by the calculations. These results indicate MED's are mostly
affected by the weakly bound effects in the nuclei being studied,
while the contribution of charge symmetry breaking is small as
discussed before. The reduction factors depend on the single
particle energies of $1s_{1/2}$ orbit. From $^{17}$F to other
nuclei, the Hamiltonians need to be changed because of the different
bindings of $1s_{1/2}$ orbit. As some nuclei being studied have no
or insufficient experimental information to obtain the single
particle energy of $1s_{1/2}$ orbit, we do not include this
nucleus-dependent effect in present work. The $1s_{1/2}$ orbits in
some nuclei are beyond the proton decay threshold. Such as, the
first $1/2^{+}$ state of $^{19}$Na, almost a purely single
$1s_{1/2}$ state, is $1.067$ MeV beyond the proton threshold. More
specific studies including nucleus-dependent and continuum effect
may be helpful in order to understand the structure of these nuclei.

Figure~\ref{21Al} shows the comparisons between data and
calculations with USD and USD* for the $A=21$ mirror pair, resulting
in a $5/2^{+}$ ground state for $^{21}$Al, which supports the
experimental assignment. An $1/2^{+}$ state is predicted. The MED is
not large enough to reverse the $5/2^{+}$ and $1/2^{+}$ states. The
one-proton separation energy in $5/2^{+}$ and $1/2^{+}$ states of
$^{21}$Al are $-1.27$ and $-2.02$~MeV in calculations with the USD*
interaction.

\begin{table}
\caption{\label{BE}Experimental and calculated binding energies (in
MeV) with the original and modified USD Hamiltonians. Data are from
Ref.~\cite{audi2012}.}
\begin{ruledtabular}
\begin{tabular}{cccccc}
  nucleus&Expt.&USD*&$|$Expt.$-$USD*$|$&USD&$|$Expt.$-$USD$|$\\
\hline
$^{18}$Ne    & 132.14    &     132.17 &    0.03 & 132.34 &  0.20  \\
$^{19}$Na    & 131.82    &     131.85 &    0.03 & 131.95 &  0.13  \\
$^{20}$Mg    & 134.48    &     134.81 &    0.33 & 134.97 &  0.49  \\
$^{21}$Al    &             &   133.54 &         & 133.61 &        \\
$^{22}$Si    &           &     135.36 &         & 135.46 &        \\
\end{tabular}
\end{ruledtabular}
\end{table}

\begin{table}
\caption{\label{reducepn}Calculated reduction factors for the six
proton-neutron TBME in which the proton $1s_{1/2}$ orbit is
involved.}
\begin{ruledtabular}
\begin{tabular}{cc}
  TBME ($\langle ij|V|kl\rangle^{pn}_{JT}$)&Reduction factor\\
\hline $\langle1s_{1/2}0d_{5/2}|V|1s_{1/2}0d_{5/2}\rangle^{pn}_{31}$&0.78\\
$\langle1s_{1/2}0d_{5/2}|V|1s_{1/2}0d_{5/2}\rangle^{pn}_{21}$&0.84\\
$\langle1s_{1/2}0d_{5/2}|V|(0d_{5/2})^{2}\rangle^{pn}_{21}$&0.87\\
$\langle1s_{1/2}0d_{5/2}|V|1s_{1/2}0d_{5/2}\rangle^{pn}_{30}$&0.81\\
$\langle1s_{1/2}0d_{5/2}|V|1s_{1/2}0d_{5/2}\rangle^{pn}_{20}$&0.80\\
$\langle1s_{1/2}0d_{5/2}|V|(0d_{5/2})^{2}\rangle^{pn}_{30}$&0.87\\
\end{tabular}
\end{ruledtabular}
\end{table}

\begin{table*}
\caption{\label{pn}Experimental and calculated binding and
excitation energies (in MeV) of $^{22-24}$Al and $^{23,24}$Si with the
original and modified USD Hamiltonians. The last column gives the
experimental excitation energies of the corresponding states in
their mirror nuclei. Data are from
Ref.~\cite{audi2012,nndc,gade2008,ichikawa2009}.}
\begin{ruledtabular}
\begin{tabular}{ccccccc}
  $spin^{parity}$ &Expt.&USD*&USD*($V^{pn}_{T=1}$)&USD*($V^{pn}_{T=1,0}$)&USD&Expt.(mirror nuclei)\\
\hline & $^{22}$Al       &       &        &        &        &$^{22}$F \\
$4^{+}$      & &149.69 	 & 149.68 &   149.60 & 149.71 &  149.74\footnotemark[1]        \\
$3^{+}$      &    &0.14     & 0.15     & 0.18   & 0.19   &0.07     \\
$2^{+}$      &    &0.64     & 0.68     & 0.76   & 0.71   &0.71     \\
\\
 & $^{23}$Al       &       &        &        &        &$^{23}$Ne \\
$5/2^{+}$      & 168.72 &168.90 & 168.88 &   168.68 & 168.88 &  168.94\footnotemark[1]        \\
$1/2^{+}$      & 0.55   &0.59     & 0.57     & 0.79   & 1.00   &1.02     \\
$3/2^{+}$      & 1.62   &1.56     & 1.56     & 1.72   & 1.77   &1.70 \\
$7/2^{+}$     & 1.77   &1.70     & 1.70     & 1.76   & 1.76   &1.82 \\
\\
 & $^{24}$Al       &       &        &        &        &$^{24}$Na\\
$4^{+}$        & 183.59 &183.72 & 183.71 &   183.40 & 183.68 & 183.77\footnotemark[1]\\
$1^{+}$        & 0.43   &0.43     & 0.43     & 0.50   & 0.45   &0.47 \\
$2^{+}$      & 0.51   &0.53     & 0.54     & 0.63   & 0.59   &0.56 \\
$1^{+}$        & 1.09   &0.87     & 0.87     & 1.06   & 1.09   &1.35 \\
\\
 &   $^{23}$Si     &       &        &        &        &$^{24}$F\\
$5/2^{+}$      &  &151.95 & 151.94 &   151.79 & 151.99  & 151.99\footnotemark[1]\\
\\
 &   $^{24}$Si     &       &        &        &        &$^{24}$Ne\\
$0^{+}$      & 172.02 &172.48 & 172.46 &   172.17 & 172.50  & 172.51\footnotemark[1]\\
$2^{+}$        & 1.88   &2.04     & 2.04     & 2.13   & 2.15   &1.98      \\
$2^{+}$       & 3.44   &3.49     & 3.46     & 3.58   & 3.74   &3.87
\end{tabular}
\end{ruledtabular}
\footnotetext[1]{The energy listed here has been modified to be
comparable with the binding energy of its mirror partner through
$E=BE(A,Z)_{expt.}+E_{C}(Z)-E_{C}(Z')$ where $E_{C}(Z)$ is the
Coulomb correction energy and $Z'$ is the proton number of its
mirror partner.}
\end{table*}

\begin{figure}
\includegraphics[scale=0.165]{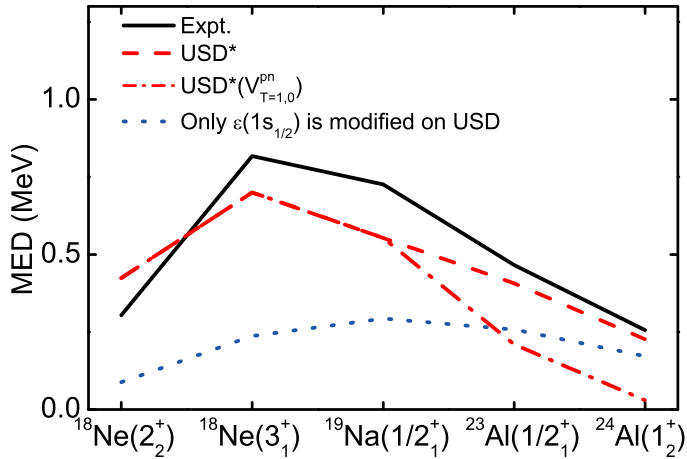}
\caption{\label{MED} (Color online) Experimental and calculated MED
for selected states in which the experimental values are larger than
$200$~keV.}
\end{figure}

\begin{table*}
\caption{\label{BGT}Comparison of experimental and calculated
$B(GT^{+})/B(GT^{-})$ values, where $B(GT^{+})$ and $B(GT^{-})$ are
the Gamow-Teller strengths for the $\beta^+$ and $\beta^-$ decays
from $^{24}$Si and $^{24}$Ne, respectively. The calculated results
are obtained with the original (for $^{24}$Ne) and modified (for
$^{24}$Si) USD, USDA and USDB Hamiltonians. Experimental values are
taken from Ref.~\cite{ichikawa2009}.}
\begin{ruledtabular}
\begin{tabular}{cccccccc}
              & &  \multicolumn{3}{c}{$\langle p,s_{1/2}|n,s_{1/2}\rangle=1.0$}     & \multicolumn{3}{c}{$\langle p,s_{1/2}|n,s_{1/2}\rangle=0.9$} \\
              &  Expt.  &     USD  &    USDA   &   USDB &  USD & USDA & USDB \\
\hline
$B(GT^{+},1^{+}_{1})/B(GT^{-},1^{+}_{1})$& 0.78(11) & 0.96 & 0.90 & 0.98 & 0.85 & 0.73 & 0.85    \\
$B(GT^{+},1^{+}_{2})/B(GT^{-},1^{+}_{2})$& 0.90(8)  & 0.88 & 0.93 & 0.85 & 0.84 & 0.87 & 0.82    \\
\end{tabular}
\end{ruledtabular}
\end{table*}

The ground-state spin of the nucleus $^{22}$Al has not yet been
determined experimentally. For its mirror nucleus $^{22}$F, the
ground state is assigned to be a $4^{+}$ state~\cite{nndc}.
Meanwhile a low-lying $3^{+}$ state has also been observed at
$71.6$~keV. The shell-model calculations can reasonably reproduce
these states. Our calculations suggest that these two states are
dominated by the coupling
$|0d^1_{5/2,t}\otimes0d_{5/2,t'}^{-1}\rangle$ where $t=n,~t'=p$ (or
vice versa) denoting the isospin of the orbits. For the $3^{+}$
state the second largest component is
$|1s^1_{1/2,t}\otimes0d_{5/2,t'}^{-1}\rangle$ which may induce a
large MED. Indeed, the MED of the $3^{+}$ state between $^{22}$F and
$^{22}$Al are as large as $57$, $69$ and $67$~keV in calculations
with the USD, USDA and USDB interactions, respectively. From these
results, the $3^{+}$ state in $^{22}$Al is predicted to be
 above the
$4^{+}$ state, which is calculated to be the ground state in the
present work, by less than $15$~keV. An analysis through the $\beta$
decay of $^{22}$Al also suggests that the ground state of $^{22}$Al
is most likely to be the $4^{+}$ state~\cite{achouri2006}.

The modified shell-model Hamiltonians can also give a good description to the binding energies of the $N=8$
isotones, $^{18}$Ne, $^{19}$Na and $^{20}$Mg, as shown in Table~\ref{BE}. In these cases only the proton-proton part of the two-body interaction, $V^{pp}$, contributes to the binding and excitation energies. The binding energy is calculated as~\cite{usdab},
\begin{equation}\label{wsho}
 BE(A,Z)=BE(A,Z)^{r}+BE(^{16}\text{O})-E_{C}(Z),
\end{equation}
where $BE(A,Z)^{r}$ and $BE(^{16}\text{O})$ denote the shell-model
energy of the nucleus $(A, Z)$ relative to the $^{16}$O core and the
experimental binding energy of the $^{16}$O core, respectively.
$E_{C}(Z)$ is the Coulomb correction energy, which is $7.45~(Z=10)$,
$11.73~(Z=11)$, $16.47~(Z=12)$, $21.48~(Z=13)$ and
$26.78~(Z=14)$~MeV~\cite{usdab}. For the nuclei investigated, the
USD interaction gives on average $0.1$~MeV better results for the
binding energies in comparison with those of USDA and USDB in both
proton- and neutron-rich side. This may be due to the fact that the
USDA and USDB interactions are built in a broader basis by including
the binding energies of many extreme neutron-rich nuclei including
$^{24}$O in the fitting besides the nuclei of concern.

For nuclei $^{22-24}$Al and $^{23-24}$Si, one needs to consider the reduction effect of the interaction matrix element of $V^{pn}$,
which also contribute to the binding and excitation energies of those nuclei, that are related to the weakly bound proton $1s_{1/2}$ orbit.
Table~\ref{reducepn} presents the
related reduction factors $V^{pn}/V^{np}$ which are evaluated by the
same method to obtain $V^{pp}/V^{nn}$.

We modified on USD* with $V^{pn}/V^{np}$ in two steps. Firstly, only
$T=1$ channel of the $V^{pn}/V^{np}$ is modified, labeled as
USD*($V^{pn}_{T=1}$). Secondly, both the $T=1$ and $T=0$ channel of
$V^{pn}/V^{np}$ are modified on USD*, labeled as
USD*($V^{pn}_{T=1,0}$). Our calculations for nuclei $^{22-24}$Al and
$^{23,24}$Si are given in Table~\ref{pn} together with experimental
data. For comparison, in the last column of the table we also give
the experimental excitation energies of the corresponding states of
the mirror partners of the nuclei of concern. It should be mentioned
that the USD interaction, where isospin symmetry is assumed, will
give the same results for the mirror partner.

Table~\ref{pn} shows that the $T=1$ channel contribute little to
both binding and excitation energies compared with USD* (on average
$14$ keV difference for these states). On the other hand, the
modification of the strongly attractive $T=0$ channel changes
significantly both binding and excitation energies compared with
USD* (on average $150$ keV difference for these states). As
indicated in Ref.~\cite{umeya2008}, the monopole channel of the
$T=0$ central force, which is strongly attractive, contributes a lot
to the binding energies in $sd$ shell nuclei, while the
contributions of two components in $T=1$ channel of the central
force are canceled to a large extent.

As shown in Table~\ref{pn}, the modification of $T=0$ channel well
reproduces the binding energy difference between the proton-rich
nuclei and their mirror partners. Regarding excitation energies (and
their MED's), the comparison to experimental data shows varying
agreement. For the pair $^{23}$Al-$^{23}$Ne,
USD*($V^{pn}$) gives comparable results to those by USD*. Regarding the pair $^{24}$Al-$^{24}$Na, we
mention that the USD cannot reproduce well the excitation energy of
the second $1^{+}$ state of $^{24}$Na. The other states also show
certain discrepancies, though to less extents. The binding energy is
better reproduced by USD*($V^{pn}_{T=1,0}$) considering the original
discrepancy between USD result and the observed value in $^{24}$Na.
For the pair $^{24}$Si-$^{24}$Ne, the overall description is
improved by the present method.

In Fig.~\ref{MED} we compare the experimental and calculated MED's
of certain states in which the experimental values are larger than
$200$~keV. It is seen that shell-model calculations with only the
shift of single particle energies taken into account are not enough
to describe the experimental MED's, while the USD* including the
modification of residual interactions can reproduce the observations.
From Table~\ref{pn} and Fig.~\ref{MED}, one can find that the modification of $T=0$ channel generally gives smaller MED's compared with observed values.
This is possibly due to the renormalization effect caused by the modification of $T=0$ channel which is not included in present study.

For heavier proton-neutron open-shell nuclei such as $^{23,24}$Al
and $^{24}$Si, additional effects may need to be considered in
future studies to give a more detailed description. One effect may
be the evolution of the single particle energies of $1s_{1/2}$ orbit
which are due to nuclear forces but not fully included in USD.

The effect discussed so far also influences the decay properties.
For example, the $B(GT)$ value of the $\beta^+$ decay from $^{24}$Si
into $^{24}$Al is smaller than that of the decay of the mirror
nucleus~\cite{ichikawa2009}. In Table~\ref{BGT}, we present the
comparison of $B(GT)$ values between the mirror nuclei, $^{24}$Si
and $^{24}$Ne. The modified USD family can describe the smaller
$B(GT)$ values of $^{24}$Si compared with those of $^{24}$Ne.

The consideration of the weakly bound effect can reduce the
$B(GT^{+})/B(GT^{-})$ value because the overlap between the radial
wave functions of the weakly bound proton and well bound neutron
$1s_{1/2}$ orbit, $\langle p,s_{1/2}|n,s_{1/2}\rangle$, is smaller
than the unity which is assumed in the conventional shell-model
calculations. As in Sec. \ref{sec:level2}, we estimate the radial
wave function of the proton $1s_{1/2}$ state by using the
Woods-Saxon potential. The depth of the potential is taken to be
$V_{0}=46.5$~MeV, while other parameters are the same as before. The
$1s_{1/2}$ orbital in $^{24}$Si is calculated to be weakly bound by
$0.1$~MeV, which is reasonable by taking into account the fact that
both the ground state of $^{25}$P and the $1/2^{+}_{1}$ state of
$^{23}$Al are unbound~\cite{nndc}. The radial wave function of the
neutron $1s_{1/2}$ state is calculated with harmonic oscillator
potential with $A=24$. The overlap between the calculated proton and
neutron radial wave functions is estimated to be $\langle
p,s_{1/2}|n,s_{1/2}\rangle=0.9$. With this value for $B(GT^{+})$,
Table~\ref{BGT} suggests that the $B(GT^{+})/B(GT^{-})$ values
obtained with the present reduction factors become sufficiently
small giving agreement with the corresponding experimental data
within errors.

\section{\label{sec:level4}Summary}

In this paper, the structure of loosely bound proton-rich nuclei
around $A=20$ are investigated within the shell model approach. We
start with several well-defined empirical shell-model Hamiltonians
constructed for this region. When applying these Hamiltonians to
nuclei in the proton-rich side, many of which would be weakly bound,
one needs to consider two important points: the shift of
single-particle energies and the reduction of the TBME. The
reduction factors of TBME are evaluated from the newly introduced NN
interaction $V_{MU}$. The large experimental MED's in $^{18}$Ne,
$^{19}$Na and $^{23}$Al are reproduced well by the modified
shell-model Hamiltonians. We predict that the $3^{+}$ state in
$^{22}$Al has an energy slightly higher than the $4^+$ ground state.
The ground state of $^{21}$Al is predicted to be $5/2^{+}$ state,
where the MED is not large enough to make the $1/2^{+}$ state lower
than the $5/2^{+}$ state.

We have also investigated the Gamow-Teller transitions for the pair
of mirror nuclei, $^{24}$Si and $^{24}$Ne. The observed
$B(GT^{+})/B(GT^{-})$ can be reproduced well by taking into account
the weakly bound nature of the proton $1s_{1/2}$ orbit.

\section*{Acknowledgement}
This work has been supported by the National Key Basic Research
Program of China under Grant No. 2013CB834400, the National Natural
Science Foundation of China Under Grant Nos. 11305272, 11235001,
11320101004, the Specialized Research Fund for the Doctoral Program
of Higher Education Under Grant No. 20130171120014, and by Japanese
MEXT Grant-in-Aid for Scientific Research (A) 20244022. C.Q.
acknowledges the supports by the Swedish Research Council (VR) under
grant Nos. 621-2010-4723 and 621-2012-3805, and the computational
support provided by the Swedish National Infrastructure for
Computing (SNIC) at PDC and NSC.

\end{document}